\documentclass[useAMS,usenatbib]{mn2e}
 
\usepackage{url}
\usepackage{graphicx}
\usepackage{amsmath}
\usepackage[usenames]{color}
\usepackage{multirow}

\title[Observing ultra-compact binaries with gravitational waves and optical telescopes.]{Prospects for observing ultra-compact binaries with space-based gravitational wave interferometers and optical telescopes.}

\author[T. B. Littenberg, S. L. Larson, G. Nelemans, N. J. Cornish]
{T. B. Littenberg$^{1,2}$\thanks{E-mail:tyson.b.littenberg@nasa.gov} 
S. L. Larson$^{3}$
G. Nelemans$^{4,5,6}$
N. J. Cornish$^{7}$\\
$^{1}$Maryland Center for Fundamental Physics, Department of Physics, University of Maryland, College Park, Maryland, 20742, USA\\
$^{2}$Gravitational Astrophysics Laboratory, NASA Goddard Spaceflight Center, 8800 Greenbelt Road, Greenbelt, Maryland, 20771, USA\\
$^{3}$Department of Physics, Utah StateUniversity, Logan, Utah, 84322, USA\\
$^{4}$Department of Astrophysics, Radboud University Nijmegen, P.O. Box 9010, 6500 GL Nijmegen, The Netherlands\\
$^{5}$Institute for Astronomy, KU Leuven, Celestijnenlaan 200D, 3001 Leuven, Belgium\\
$^{6}$Nikhef, Science Park 105, 1098 XG Amsterdam, The Netherlands\\
$^{7}$Department of Physics, Montana State University, Bozeman, Montana, 59717, USA}
\begin{document}

\date{Accepted --- Received ---; in original form ---}

\pagerange{\pageref{firstpage}--\pageref{lastpage}} \pubyear{2012}

\maketitle

\label{firstpage}

\begin{abstract}
Space-based gravitational wave interferometers are sensitive to the
galactic population of ultra-compact binaries.  An important subset of
the ultra-compact binary population are those stars that can be
individually resolved by both gravitational wave interferometers and
electromagnetic telescopes.  The aim of this paper is to quantify the
multi-messenger potential of space-based interferometers with
arm-lengths between 1 and 5 Gm.  The Fisher Information Matrix is used
to estimate the number of binaries from a model of the Milky Way which
are localized on the sky by the gravitational wave detector to within
1 and 10 deg$^2$ and bright enough to be detected by a magnitude limited
survey.  We find, depending on the choice of GW detector characteristics,
limiting magnitude, and observing strategy, that up to several hundred
gravitational wave sources could be detected in electromagnetic
follow-up observations.
\end{abstract}


\section{Introduction}\label{sec.Intro}

A variety of detector concepts for space-based gravitational wave
interferometers have been proposed, the most well studied concept
being LISA\citep{LISA}.  It was understood early on that the most
numerous source class radiating in the band covered by LISA-like
detectors will be the galactic population of ultra-compact binaries
(UCBs) comprised of pairs of stellar remnants: white dwarfs, neutron
stars or black holes.  The gravitational radiation from these UCBs will 
be the dominant signal in the frequency band covered by LISA-like detectors.

Early estimates of the composite signal from the UCBs
\citep{Evans1987, HBW,HB1997} demonstrated that the signals of the vast majority
of the galactic binaries will overlap and be unresolvable from one
another, forming a limiting foreground (or ``confusion noise'') for
space-based gravitational wave detectors.  Later studies based on
population synthesis \citep{Nelemans, Benacquista, Edlund2005, TRC,
RBBL} have borne this expectation out.  Detailed data analysis studies
have shown that $\sim 10^{4}$ individual binaries could be resolved
out of the foreground by a gravitational wave observatory like LISA
\citep{TRC, Crowder2007, Littenberg2011, Nissanke2012}.

A subset of the resolvable binaries will be detectable
electromagnetically.  The purpose of this work is to assess the
multi-messenger potential for different space-based detectors spanning
the trade-space of future mission designs.  This builds off previous
work \citep{Cooray2003, Nelemans2006, Nelemans2009} demonstrating the
feasibility of follow-up observations for high-frequency UCB sources.
We estimate the total number of multi-messenger sources by beginning
with a population synthesis model of the galaxy \citep{Nelemans2004},
complete with optical magnitudes.  From this we produce a magnitude
limited source catalog, then estimate how well each system will be
localized on the sky by different gravitational wave detector
configurations.  Using hundreds of Monte Carlo realizations over the
spatial distribution of the galaxy and the UCB orientations, we find
tens to hundreds of sources that can be observed both
electromagnetically and gravitationally.

The information encoded about the UCBs in each of the two spectrums 
is highly complementary, enabling tests of general relativity, 
full measurement of the physical parameters enabling constraints on 
binary synthesis channels, and new methods of probing the close 
interaction dynamics of the compact stars \citep{Cutler2003, Stroeer2005}.

\section{Detectors}\label{sec.detector}
For a gravitational wave observatory, the limiting sensitivity as a
function of frequency is dominated at low frequencies by acceleration
noise $S_{a}$, while the ``floor'', where the detector is most
sensitive, is dominated by position measurement noise $S_{x}$.  Table
\ref{tbl.detectors} contains the parameters used for the detector
configurations in this study.  These parameters can be used to compute
the noise power spectral density
\begin{eqnarray}
   S_n(f) &=& S_{\rm{gal}}(f) + (4/3)\sin^2{u} \left[(2+\cos{ u} ) 
   S_{x} \right. \nonumber \\
   &+& \left. 2(3 + 2\cos{u}+\cos{2u} ) S_{a}/(2\pi f)^4\right] ,
   \label{eqn.Sn}
  \end{eqnarray}
where $u=2\pi f\ell/c$ and $S_{\rm{gal}}$ is the contribution to the
instrument noise from the unresolved UCB foreground~\citep{TRC}.

The configurations we will highlight correspond to the classic LISA
design ($\ell = 5$ Gm), as well as two shorter arm-length
configurations ($\ell = 2$ Gm and $\ell = 1$ Gm) in order to cover a
variety of plausible mission configurations.  
The 1 Gm configuration is similar to the eLISA mission being considered 
by the European Space Agency~\citep{eLISA}.  We use an observation
time of two years for each configuration.  

\begin{figure}
  \centering
 \includegraphics[angle=270, width=0.5\textwidth]{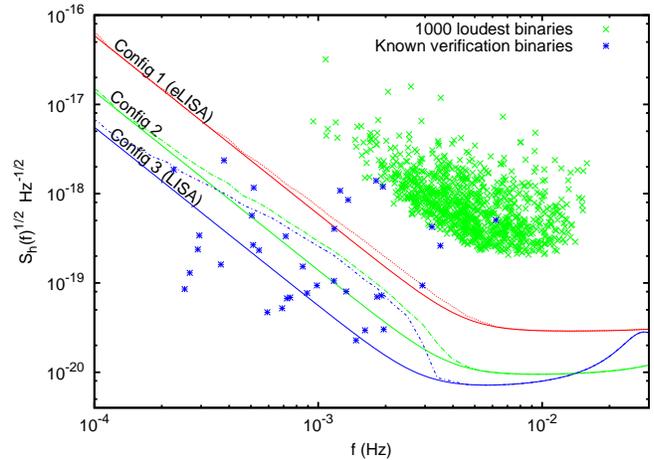} 
  \caption{{\small Sensitivity curve for each of the detector
  configurations in Table \ref{tbl.detectors}.  The solid lines show the sensitivity set by the measurment noise while the dashed curves include an estimate of the UCB confusion-limited foreground.  Over-plotted are the
  brightest UCBs in our simulated catalog (green crosses), and the known verification binaries (blue stars)}}
  \label{fig.detectorPSD}
\end{figure}

This suite of detectors
provides a broad palette to illustrate the observational capabilities
of these instruments with regards to the UCBs.  A classic depiction of
the performance for these interferometers is a plot of the average
sensitivity curve in strain spectral density versus frequency
\citep{SCG}, as shown in Fig.~\ref{fig.detectorPSD}. 
The eLISA concept is the only 
one which uses a 4-link configuration.  The doppler ranging between each spacecraft in the 
constellation is accomplished using two laser links.  Thus the 4-link design is a single-vertex 
interferometer, while the 6-link designs allow for three (coupled) interferometers.  This difference accounts for an additional improvement in the 6-link sensitivity curves by a factor of $\sim\sqrt{2}$ at frequencies where the UCBs are found. 

\begin{table}
    \centering
    \begin{tabular}{rcccc}
	 \hline \hline Config. & $\ell$  &
	 $\sqrt{S_{a}}$  &
	 $\sqrt{S_{x}}$ & Links\\
	 &(m)&(m/s$^{2}/\sqrt{\rm Hz}$)&(m/$\sqrt{\rm Hz}$)&\\
      \hline
   1 & $1 \times 10^{9}$ &   $4.5\times10^{-15}$ &  $11\times10^{-12}$ & 4\\
   2 & $2 \times 10^{9}$ &   $3.0\times10^{-15}$ &  $10\times10^{-12}$ & 6\\
   3 & $5 \times 10^{9}$ &   $3.0\times10^{-15}$ &  $18\times10^{-12}$ & 6\\
      \hline
  \end{tabular}
	\caption{{\small Gravitational wave detector configurations used in this study.  Configuration 1 corresponds to eLISA.  Configuration 5 is the classic LISA design.  All simulations were for two year mission lifetimes.}}

  \label{tbl.detectors}
\end{table}

\section{Discovering new verification binaries}\label{sec.discovery}
The focus of this work is to study the population of detectable UCBs
in the context of multi-messenger astronomy.  We will focus on the
sources detected via GWs which could potentially be identified
electromagnetically.  There is a separate class of UCBs, the
``verification binaries,'' which are known low-frequency GW sources
with AM CVn serving as the archetype.  There are $\sim30$ known
verification binaries, $\sim 5-10$ of which could be identified by the GW detectors considered here, with sources still being discovered
\citep{NelemansWiki, Roelofs2007, Brown2011}.  
This study does not include the known verification binaries in the galaxy catalogs.  Furthermore, many of the AM CVn systems would \emph{not be localized well enough by the GW measurement alone} to warrant simple electromagnetic follow-up observations.  

\subsection{Binary selection}\label{sub.BinarySelection}

The UCB population model is essentially identical to that found in \citet{Nelemans2004}, so the positions and ages of the systems
are based on the Boissier \& Prantzos (1999) Galactic model.  We use
the white dwarf cooling tracks based on Hansen (1999) as shown in the
Appendix of \citet{Nelemans2004}.  We convert the luminosities to
V-band magnitudes using zero-temperature white dwarf radii and simple
bolometric corrections based on the effective temperature.  This
should suffice for this initial estimate of the potentially detectable
population, but can be improved using detailed WD cooling models in
the future.  We determine the absorption as in \citet{Nelemans2004}
based on the Sandage (1972) model, but correcting for the fact that
the dust is more concentrated than the stars, so we use 120pc as scale
height for the absorption. 

To construct the magnitude limited catalog, we begin with the entire binary population in the synthesized galaxy.
The limiting apparent magnitude of a telescope is a function of the
aperture $D$, the exposure time $t$, and the properties of the
detector used for imaging and photometry
\citep{Schaeffer1990,Howell1989}.  A rudimentary fit to the limiting
magnitude $m$ using a telescope of aperture $D$ (in m) and for
exposure time $t$ (in seconds) is given by $m = 19.6\;
D^{0.073}\;t^{0.025}$.  Using commercial CCD detectors, a $D = 0.5$ m
telescope will reach a photometric magnitude $m \simeq 21$ in $t\sim
75$ s, where as a $D = 1.0$ m telescope will reach the same magnitude
in $t\sim 20$ s.  This paper examines the role of small to large
aperture telescopes by examining a broad range of limiting magnitudes;
lower bounds of $m = 18-24$ were chosen as the electromagnetic cutoff.
All sky survey instruments such as LSST could further improve the number of candidates.  The single exposure limit for LSST is
expected to be $m \simeq 24$, whereas the magnitude limit of the final
stacked image is expected to be around $m \simeq 27$
\citep{Ivezik2011}.

\subsection{Gravitational wave detector response}

From the magnitude-limited catalog, we determine the number of
``bright'' UCBs that will be well measured by the GW detector.  To do
so we must first estimate the confusion noise for each configuration.
The instrument response to the galactic foreground is constructed by
generating and co-adding waveforms for each source in the full
simulated galaxy catalogue using the fast-slow decomposition in
\citet{Cornish2007}.  The confusion noise, $S_{gal}$, is empirically
determined from the simulated data by iteratively removing sources
brighter than a running estimate of the background.  This procedure is
first discussed in \citet{TRC} with an improved implementation 
used here as in \citet{Nissanke2012}.

With the confusion noise incorporated into the detector sensitivity
curves, we determine how well the GW detector can measure the source
parameters of a UCB waveform, using the well-worn
Fisher Information Matrix $\Gamma_{ij}$ \citep{Cutler1994}, the inverse of which
approximates the covariance matrix.  There is no
shortage of literature highlighting short-comings of the Fisher to
approximate GW parameter errors e.g., \cite{Vallisneri2008}.  However,
given the scope of the problem we are addressing (hundreds of Monte
Carlo's of thousands of detectable binaries) more rigorous parameter
estimation studies would be impractical (recently \citet{Vallisneri2011} has proposed a way around this dilemma).  On the other hand, the UCBs in
which we are most interested -- those that can be well localized on the
sky -- have atypically high signal to noise ratio, where the Fisher
provides a good estimate of the true parameter errors
\citep{Crowder2007}.


For a binary to be considered ``well localized'' we require that the 63\% confidence 
interval of the sky-location posterior distribution function subtends an area on the celestial sphere below some threshold $d\Omega$.  To bracket the capabilities of ground-based optical telescopes, we perform the analysis with $d\Omega\leq1$ and $\leq10$ deg$^2$.  
We estimate the area of the sky-location error ellipse using the full covariance matrix 
found by inverting $\Gamma_{ij}$ \citep{Lang2008}.

\begin{figure*}
  \centering
 \includegraphics[width=\textwidth]{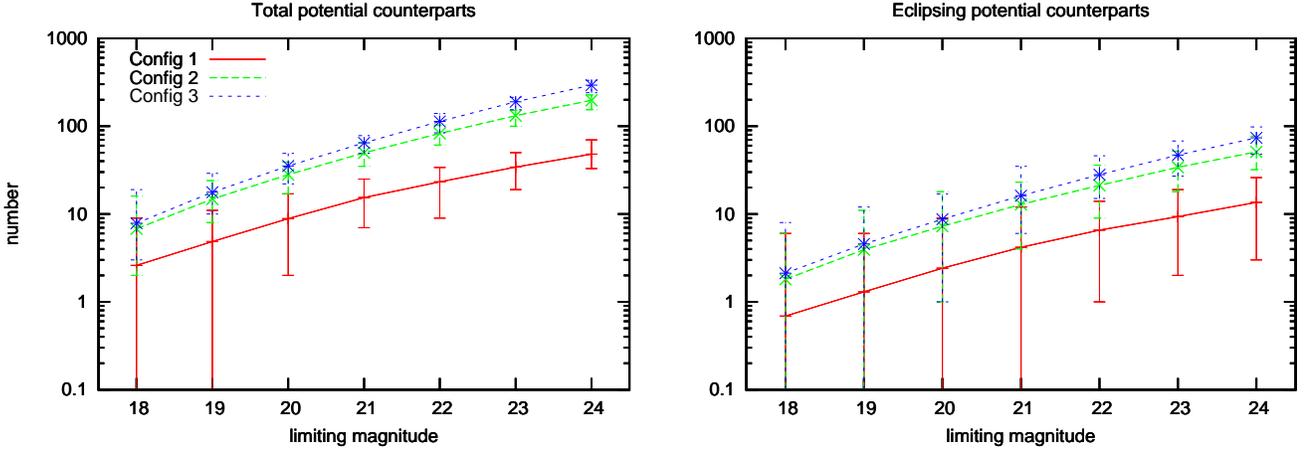} 
 \caption{{\small Number of binaries with sky-location resolved
 to within 1 deg$^{2}$ for each configuration as a function of
 limiting magnitudes.  The left panel shows the total number of candidates, the right panel shows the subset of eclipsing binaries.
 The error bars represent the full range after Monte Carlo'ing over
 the location and orientation of each UCB system.  Even the
 modest detection abilities (magnitudes $m \sim 19$) of small
 aperture telescopes can yield several electromagnetic counterparts;
 larger telescopes with deeper magnitude grasp will have
 significantly more sources that can be surveyed. }}
 \label{fig.numberSources1}
\end{figure*}

\begin{figure*}
  \centering
 \includegraphics[width=\textwidth]{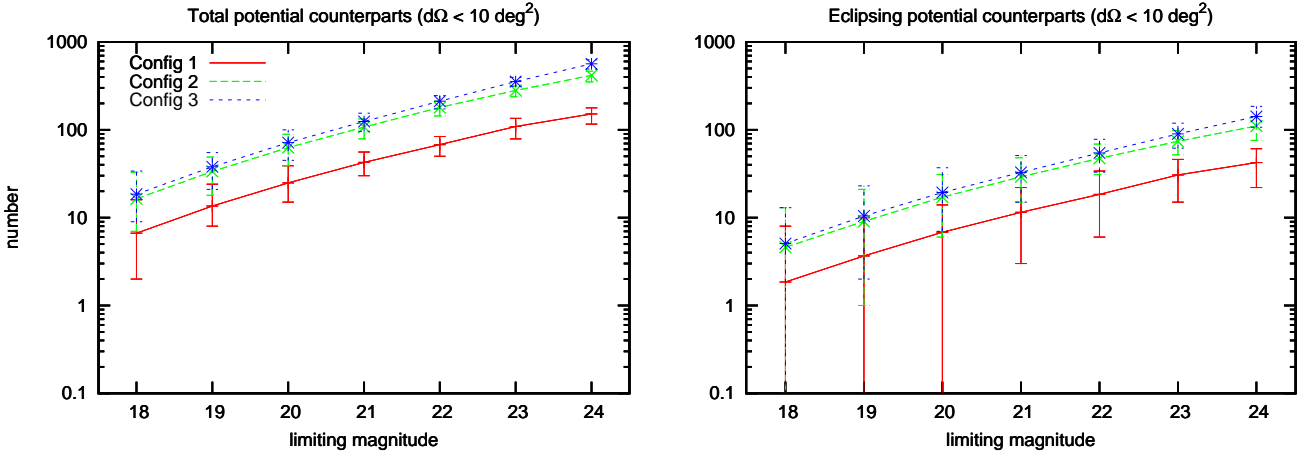} 
 \caption{{\small Same as Figure~\ref{fig.numberSources1}, except here we use an angular resolution threshold for the GW detector of $d\Omega\leq10$ deg$^2$. }}
 \label{fig.numberSources10}
\end{figure*}

The number of well localized, bright binaries is computed for hundreds of 
realizations where we Monte Carlo over the orientation
of each binary, as well their location within the Galaxy.
For the orientation, we draw the inclination $\iota$ from a uniform distribution $\cos\iota = {\rm U}[-1,1]$, and the polarization angle $\psi$ and initial phase $\varphi$ from U$[0,2\pi]$.
We find up to a several hundred GW sources will be viable
candidates for electromagnetic follow-up searches, depending on the
depth of EM survey and the GW detector characteristics (See the left-hand
panel of Figs.~\ref{fig.numberSources1}~and~\ref{fig.numberSources10}). 

\begin{table*}
         \centering

    \begin{tabular}{c  rcc  rcc  rcc}
	 \hline \hline 
	 
	 & Full galaxy &&& $m\leq20$ &&& $m\leq24$ &&\\
	Config & \multicolumn{1}{c}{$\bar{N}$}& $|b|>20^{\circ}$ & $\frac{\Delta d_L}{ d_L} < 20\%$ &  \multicolumn{1}{c}{$\bar{N}$}& $|b|>20^{\circ}$ & $\frac{\Delta d_L}{ d_L} < 20\%$&  \multicolumn{1}{c}{$\bar{N}$} & $|b|>20^{\circ}$ & $\frac{\Delta d_L}{ d_L} < 20\%$\\ 
      	\hline


	1  & 1700 (370) & 0.02 (0.03)& 0.60 (0.93) & 24 (9) & 0.44 (0.49) & 0.16 (0.43) & 150 (48) & 0.22 (0.25) & 0.26 (0.67)\\ 
	2  & 7500 (2400) & 0.01 (0.01) & 0.34 (0.75) & 62 (28) & 0.48 (0.48)  & 0.08 (0.17) & 444 (197) & 0.22 (0.22) & 0.16 (0.34)\\ 
	3  & 12000 (6100) & 0.01 (0.01) & 0.27 (0.60) & 71 (35) & 0.48 (0.40) & 0.07 (0.15) & 563 (293) & 0.22 (0.21) & 0.14 (0.26)\\ 
		

	 \hline
  	\end{tabular}
  \caption{{\small Multi-messenger candidates will be a minority of the spatially well-resolved GW signals.  Here we enumerate the fraction of binaries that will make good candidates for electromagnetic follow-up observations.  Plain numbers in the table correspond to the $d\Omega\leq 10$ deg$^2$, while those in parenthesis correspond to the 1 deg$^2$ threshold.  We tabulate the average number of binaries $\bar{N}$ that meet the sky-resolution requirements (column 2), and then the fraction which are significantly above the disk ($|b| > 20^{\circ}$ -- column 3), or have $d_L$ measured to within 20\% (column 4), the idea being that near-by binaries are more likely to be optically detectable -- proximity can be inferred by $|b|$ and/or determined through $d_L$.  The columns then repeat for the $m\leq20$ and $m\leq24$ magnitude-limited catalogs.}}
  \label{tbl.strategy}
\end{table*}

\subsection{EM detection strategies}
We now consider how to select candidates for follow-up observations
from the full GW catalog.  Pointing telescopes at all of the GW
sources localized within the adopted threshold would be inefficient, as we find
between $10^3$ and $10^4$ GW sources in the full catalog will meet the $d\Omega\leq10$ deg$^2$ threshold, while $\la 10\%$ are likely to be brighter than $m=24$, and only $\la1\%$ pass the $m\leq20$ cut.

Additional considerations need to be made to increase the efficiency of follow-up observing campaigns.  We illustrate two simple ways to
isolate the GW sources that may be electromagnetically observable.  
These suggestions are supported by calculations shown in Table~\ref{tbl.strategy}.

First, the large majority of UCB sources are confined within the
galactic plane.  Conversely, the magnitude limited catalogs sample the
local galaxy, which is much more uniformly distributed on the
celestial sphere.  Therefore, as a rough cut on the GW catalog, any
binaries that are well localized but out of the galactic plane are
good candidates.  These are additionally attractive sources, as there
will be less optical background and extinction against which the
observing campaign will have to compete.
We find between $\sim 20\%$ and $\sim50\%$ of the well-localized binaries in the 
20$^{\rm th}$ to 24$^{\rm th}$ magnitude-limited catalogs have galactic latitudes $|b| \geq 20^{\circ}$, while that fraction is reduced to $\sim1\%$ for the full GW catalog.  A uniform distribution of stars on the celestial sphere would have 66\% of the stars with $|b| \geq 20^{\circ}$ 

The other strategy for identifying optical counterparts relies on
estimates of the distance to the galactic binary.  Typical UCB sources
will undergo very little evolution of their orbital period during a
space-borne GW detector's lifetime.  Without measurement of the rate
of change of the gravitational wave frequency $\dot{f}$ the GW
observation only constrains the overall amplitude of the signal
without decoupling the chirp-mass and the luminosity distance $d_L$
\citep{Schutz1986, StroeerVecchio2006}.  
For $\sim10 - 20\%$ of the multi-messenger sources we sufficiently constrain $\dot{f}$ and $\mathcal{A}$ to measure $d_L$ to within 20\%, but astrophysical effects such as tides may impact the orbital evolution and thus bias the distance estimate.  For the remaining systems in the GW catalog, we can use reasonable priors on the
mass and mass ratio of white dwarf binaries to put meaningful constraints on $d_L$ from the amplitude measurement alone.

Using only the amplitude, frequency and priors on the masses
constructed from the population synthesis simulation, we find that the
distribution of the most likely (ML) luminosity distances $dL_{\rm
ML}$ is strongly peaked between 0 and 8 kpc -- the distance to the
galactic center -- for the magnitude limited catalogs.  The
$dL_{\rm ML}$ distributions for the full well-localized catalog with no
magnitude cut is more uniformly distributed over a larger range.

Our final consideration pertains to the expected optical light curves for UCB systems identified in the GW catalog.
The population synthesis galaxy in our study is restricted to detached white dwarf
binaries, as opposed to interacting AM CVn systems.  Without mass
transferring from one star to the other in the binary, photometric
variability is not guaranteed.  The systems in the GW catalog that are
best constrained are typically those at the high-frequency end of the
population.  This is to our advantage, because the shorter period
binaries have a higher probability of eclipsing one another during an
orbital cycle.

We can put an additional cut on our EM/GW catalog by requiring the
binaries to be eclipsing.  (See the right-hand panel of
Figs.~\ref{fig.numberSources1}~and~\ref{fig.numberSources10}).  From simple geometrical arguments
\citep{Cooray2003} the minimum inclination angle with respect
to our line of sight that will produce eclipsing light curves is
\begin{equation}
   \cos(\iota_{\rm min})\sim0.3(f/3.5\ {\rm mHz})^{2/3}
\end{equation}
assuming all binaries have mass $M_{\rm total}\sim 0.5\ {\rm
M}_{\odot}$ and radius $R_{\rm WD}\sim10^4\ {\rm km}$.  If we only
consider binaries in the multimessenger catalog with inclination angle
less than $\iota_{\rm min}$, we reduce the total number of candidates
by a factor of $\sim 3$.  Nevertheless, we still find upwards of $\sim 100$
candidates for the large GW detector configurations and deep, wide field, optical
surveys.  Requiring eclipsing light curves significantly degrades
the multimessenger potential for the 1 Gm configuration using
catalogs limited to 20$^{\rm th}$ magnitude and dimmer -- such
EM follow-up surveys could come up empty.

\section{Discussion}\label{sec.Discussion}

We conclude that space-based gravitational wave detectors will be
useful observatories for discovering new UCBs in the galaxy that could
be observed electromagnetically, though deep, wide field,  optical surveys may be
required to produce large catalogs.  We reach this verdict by
considering a range of plausible near-future space-based gravitational
wave detector concepts, and assess their measurement capabilities for
magnitude limited catalogs of UCBs.  Magnitudes for the constituents
of each binary were derived from the population synthesis simulations, and the gravitational wave measurement capabilities were
estimated using the Fisher Information Matrix.  Any UCBs that were
brighter than our chosen magnitude limits (18-24) and located on the
sky by the gravitational wave detector to within angular resolution $d\Omega$ were
considered multi-messenger candidates.  We estimated the multi-messenger catalog sizes for both $d\Omega \leq 1$ and 10 deg$^2$.

At the pessimistic end, we consider magnitude 18 limited catalogs, and
single-vertex interferometers with 1 Gm arm-lengths.  The best
scenario considered the classic LISA design and an optical telescope
limited at $24^{\rm th}$ magnitude.  The number of multi-messenger
candidates was anywhere from a few to several hundreds
over that range of detector capabilities.  If we put on the additional
constraint that the sources must be eclipsing to allow for
electromagnetic observation the counts were reduced by a factor of
$\sim 3$.

While most of the known verification binaries are AM CVn type stars,
 our study only considered detached white dwarf binaries, thus
providing a very complimentary catalog of UCB multi-messenger systems.

This work considered a conservative approach to finding
multi-messenger UCBs, with competing criteria that strongly affect the
expected population of systems detectable in both spectrums.
Electromagnetic detections are most strongly affected by the magnitude
limit of the detection survey, a function of telescope aperture.  By
contrast, the gravitational wave detection catalogs of UCBs are
expected to have thousands of systems in them; most will be too faint
to be detectable by any electromagnetic survey.  However the
gravitational wave localization criterion is a strong constraint on the multi-messenger catalog.
We find that wide-field surveys ($d\Omega\leq10$ deg$^2$) yield more candidates than more narrow fields of view ($d\Omega\leq1$ deg$^{2}$) by 50-100\% for the full catalogs, and by a factor of 2-4 for the eclipsing binaries.

We have estimated the number of UCB multi-messenger candidates without considering what could be done with joint GW and EM observations.  Our follow-on effort will consider the science yield from joint observations of both the known verification binaries -- mostly mass-transferring systems -- and the close, detached binaries that will be discovered by space-borne gravitational wave detectors.

\section*{Acknowledgments} This work was supported by NASA Grants 08- ATFP08-0126 (TBL),  NNX12AG30G (SLL), and  NNX10AH15G (NJC).  GN acknowledges support from the Dutch Foundation 
for Fundamental Research on Matter (FOM).

\bsp

\label{lastpage}

\end{document}